\documentclass[12pt]{article}
\input epsf.sty
\usepackage{hyperref}

\usepackage{hyperref}
\usepackage{amssymb}
\usepackage{amsmath}
\usepackage[nosort]{cite}
\usepackage[normalem]{ulem}

\newcommand{\be}{\begin{equation}}
\newcommand{\ee}{\end{equation}}
\newcommand{\ben}{\begin{eqnarray}\displaystyle}
\newcommand{\een}{\end{eqnarray}}

\usepackage{amsmath}
\usepackage{amsfonts}
\usepackage{amssymb}
\usepackage{cite}
\usepackage{graphicx}

\def\sqr#1#2{{\vcenter{\vbox{\hrule height.#2pt
         \hbox{\vrule width.#2pt height#1pt \kern#1pt
            \vrule width.#2pt}
         \hrule height.#2pt}}}}

\usepackage{hyperref}
\hypersetup {
	colorlinks=true,
	linkcolor=blue,
	linktocpage=true,
	citecolor=blue,
	urlcolor=blue
}
\usepackage{geometry}

\usepackage{graphicx}

\usepackage{amsmath}
\usepackage{amsfonts}
\usepackage{amssymb}
\usepackage{cite}
\usepackage{graphicx}

\usepackage{setspace}
\numberwithin{equation}{section}

\begin{document}
\begin{center}
{
\Large{\bf  The Averaged Null Energy Condition  and \\ \vspace{2mm} the Black Hole Interior in String Theory}}

\vspace{10mm}


\textit{ Karinne Attali and Nissan Itzhaki }
\break

Physics Department, Tel-Aviv University, \\
Ramat-Aviv, 69978, Israel \\
{\it karinnea@mail.tau.ac.il,  nitzhaki@post.tau.ac.il}


\end{center}

\vspace{10mm}


\begin{abstract}

Recently it was shown that folded strings are spontaneously created behind the horizon of the $SL(2,\mathbb{R})_k/U(1)$ black hole. Here we show that these folded strings violate the averaged null energy condition {\it macroscopically}. We discuss possible consequences  of this observation on black hole physics in general, and on the information puzzle in particular. 

\end{abstract}

\newpage

\section{Introduction}

The Averaged Null Energy Condition (ANEC)
 is the statement that for any state, $|\Phi\rangle $, in the Hilbert space 
\be\label{anec}
\langle \Phi |\int du T_{uu} |\Phi \rangle \geq 0,
\ee
where $u$ is a null direction. The ANEC plays an important role in theoretical physics. In 
 Quantum Field Theory (QFT), where it can also be proved using various methods \cite{Kelly:2014mra,Faulkner:2016mzt,Hartman:2016lgu}, it leads to results that are not sensitive to the details of the theory.  For example, it can be used to bound  $a/c$ in 4D conformal field theories \cite{Hofman:2008ar} (for some  extensions see e.g. \cite{Hartman:2016lgu,Komargodski:2016gci,Cordova:2017zej,Delacretaz:2018cfk}).

In General Relativity (GR) the ANEC is used to establish various fundamental properties of Black Holes (BHs) and Cosmology. For example, the ANEC is sufficient \cite{Borde:1987qr} to prove the geodesic focusing that leads to the BH area increase theorem \cite{ath}. It is of fundamental importance to understand the extent to which the ANEC holds in quantum gravity. In situations where the curvature is small, such as the horizons of large BHs, we expect the ANEC to hold up to corrections that go to zero with the curvature. The BH radiation \cite{Hawking:1974sw} illustrates this neatly as, for a large BH, it  violates the ANEC by a small amount  that decreases  the BH area.  

String theory is not expected to modify this,  at least when the string coupling constant, $g_s$, is small. When the curvature and $g_s$ are small  string theory can be viewed as a collection of  (infinitely) many fields, each of which satisfies the ANEC up to small curvature corrections. 

The goal of this paper is to challenge this expectation. We show that behind the horizon of a large $SL(2,\mathbb{R})_k/U(1)$ BH there are, ontop of the standard stringy excitations that respect the ANEC, folded strings that are spontaneously created and violate the ANEC. The violation of the ANEC by the folded string is macroscopic and, in particular, it does not vanish with the size of the BH. 

The paper is organized as follows. In the next section we show that there are folded string solutions in the time-like linear dilaton background, that are spontaneously  created and violate the ANEC at macroscopic scales. In section 3 we take the slope to the time-like linear dilaton to $0^{+}$ and show that in this limit the folded string respects a new symmetry. 
In section 4 we argue that folded strings are spontaneously  created and violate the ANEC at macroscopic scales behind the horizon of a 
large $SL(2,\mathbb{R})_k/U(1)$ BH. In section 5 we recall the proofs of the ANEC in the context of QFT and emphasise the tension between these proofs and  the folded strings. This tension, as well as other arguments such as the FZZ duality \cite{fzz,Kazakov:2000pm}, motivates us to conjecture, in section 6, that the number of folded strings that are created behind the horizon is such that their backreaction prevents information from falling to the BH.  We summaries in section 7.

\section{The ANEC and time-like linear dilaton}

In this section we discuss a simple  stringy background in which  folded strings are spontaneously  created and  violate the ANEC macroscopically. 
We consider a time-like  linear dilaton background 
\be\label{bac}
ds^2=-(dX^0)^2+(dX^1)^2,~~~\Phi=\Phi_0+Q X^0.
\ee
Our interest here is only with classical aspects of the theory.  Hence the fact that quantum mechanically this background, by itself, is not consistent plays no role here.

The case where $Q$ is negative and large was studied quite a bit in the past (see e.g. \cite{Hellerman:2006nx,Aharony:2006ra} and references therein)  as a toy model for stringy cosmology (with a singularity in the past). As can be expected in this case the linear dilaton has a dramatic effect on the physics.

Our motivation is to study the region behind the horizon of a large BH. With this motivation in mind, we consider 
 $Q$  positive, so  the singularity is in the future, and small, which is the case in the $SL(2,\mathbb{R})_k/U(1)$ BH with large $k$.   
 Since the slope of the dilaton is small, it is natural to suspect  that the linear dilaton has a small effect. It is true that for any small $Q$ the string coupling constant blows up eventually in the future and the perturbative stringy description breaks down, but with the help of $\Phi_0$ we can push this singularity arbitrarily far into the future. There is, therefore, no reason to expect any surprises  when $Q$ is small. In particular, we do expect the ANEC to hold.    

At first this seems to be the case. The  linear dilaton modifies the Virasoro constraints in the following way 
\be\label{vvc}
(\partial_{+}X^1)^2-(\partial_{+}X^0)^2-Q\partial^2_{+}X^0=0,~~~~~~~(\partial_{-}X^1)^2-(\partial_{-}X^0)^2-Q\partial^2_{-}X^0=0,
\ee
which shifts the conformal dimension  of perturbative states with energy $E$  from $-E^2/4$ to $-E(E+Q)/4$.
This implies that, as can be expected from (\ref{bac}), the linear dilaton affects,  the perturbative physics, at length scales of the order $1/Q$, that are much larger than the string scale.

Recently, however, it was pointed out that (\ref{vvc}) has another, more surprising, effect. It generates smooth classical solutions to the equation of motion and Virasoro constraint\footnote{This observation is a trivial analytic continuation of a solution found some time ago in the context of space-like linear dilaton \cite{Maldacena:2005hi}. Its physical interpretation, however,  is quite different.}. The solution takes the form 

\be\label{s}
X^1=\sigma,~~~~X^0=x^0+Q \log\left( \frac12\left( \cosh\left(\frac{\sigma-x^1}{Q}\right)+\cosh\left(\frac{\tau}{Q}\right)\right)\right),
\ee 
where the range of both $\tau$ and $\sigma$ in this solution is $-\infty$ to $\infty$.  Eq. (\ref{s}) 
 describes a closed folded string that is spontaneously created at $X^0=x^0$ and $X^1=x^1$. The points where the string folds, $\tau=0$, are moving faster than light. At the creation point the  speed is infinite and it  approaches the speed of light at later times (see figure 1). 

\begin{figure}
\begin{center}
\includegraphics[scale=0.40]{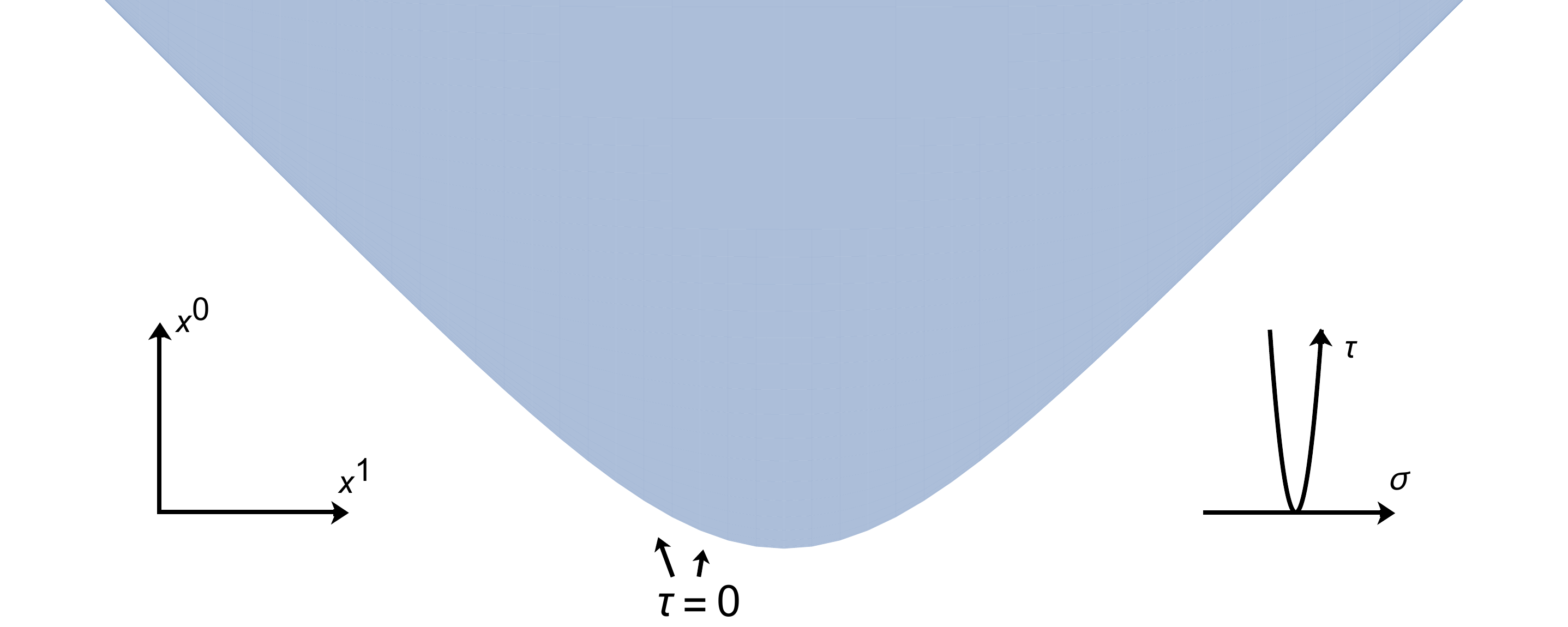}\vspace{-2mm}
\caption{The folded string configuration a 	in time-like linear dilaton background. At a certain point in space-time a closed folded string is created. The points where the string folds, $\tau=0$, travel faster than light.}
\label{SBHpen}
\end{center}
\end{figure}\label{fh1}

A useful toy model that describes the physics of the points where the string folds might be the following. Suppose that there is a field with the following kinetic term in the Largangian
\be\label{toy}
\left(1- QE\right) (E^2-P^2),
\ee
where $E$ is the energy and $P$ the momentum, and we still take $Q$ to be positive and small. Since $Q$ is small, even for stringy energies $E\sim 1$, the kinetic term is standard. However, for  $E>1/Q$ the kinetic term  flips sign and ghost condensation \cite{ArkaniHamed:2003uy} takes place.  

An important feature of the folded string  solution (\ref{s}), that appears also in the toy model above, and is crucial for its generalizations to other backgrounds, is that the length scale associated with its creation is $Q$,  which is much smaller than the curvature and string scale. In other words, the non-perturbative solution introduces a new scale to the problem that is much shorter than the perturbative scales. 

Thus,  folded strings are spontaneously created when the local inertial frame is described by a time-like linear dilaton with string coupling that grows into the future. Namely when 
\be\label{op}
(\nabla \Phi)^2<0,~~~~\mbox{and}~~~~\partial_{0}\Phi>0,
\ee
where $X^0$ is the time direction in the local inetrial frame.
 In  section 4 we elaborate more on this in the context of the region behind the horizon of the $SL(2,\mathbb{R})_k/U(1)$ BH.

Next we  show that the folded string  solution (\ref{s}) violates the ANEC. To this end we need to calculate the energy-momentum tensor associated with it. The energy momentum tensor of a classical  string  is given by
\begin{equation}
 T^{\mu\nu}(X) = \frac{1}{2 \pi \alpha^\prime}\int \,d\sigma\,d\tau\left(\dot{X}^{\mu}\dot{X}^{\nu}-X'^{\mu}X'^{\nu}\right)\delta(x-X(\sigma,\tau))\label{eq:energy momentum def}
\end{equation}
Plugging (\ref{s}) into (\ref{eq:energy momentum def}) we get 
\begin{align}
T^{00}(X^0,X^1) &= \frac{1}{2 \pi \alpha^\prime} \frac{-4\cosh\left(\frac{X^1 -x^1}{Q}\right)+ 4e^{\frac{X^0-x^0}{Q}}}{\sqrt{-1+ \left(2e^{\frac{X^0-x^0}{Q}}-\cosh\left(\frac{X^1 -x^1}{Q}\right)\right)^2}}\,\Theta\left(f(X^0,X^1)\right)\nonumber \\ 
T^{11}(X^0,X^1)&= \frac{1}{2 \pi \alpha^\prime} \frac{-4e^{\frac{X^0-x^0}{Q}}}{\sqrt{-1+ \left(2e^{\frac{X^0-x^0}{Q}}-\cosh\left(\frac{X^1 -x^1}{Q}\right)\right)^2}}\,\Theta\left(f(X^0,X^1)\right)\\
T^{01}(X^0,X^1)&= \frac{1}{2 \pi \alpha^\prime} \frac{-2\sinh\left(\frac{X^1 -x^1}{Q}\right)}{\sqrt{-1+ \left(2e^{\frac{X^0-x^0}{Q}}-\cosh\left(\frac{X^1 -x^1}{Q}\right)\right)^2}}\,\Theta\left(f(X^0,X^1)\right)\nonumber 
\end{align}
Where 
\be 
f(X^0,X^1)\equiv 2e^{\frac{X^0-x^0}{Q}}-\cosh\left(\frac{X^1 -x^1}{Q}\right)-1,
\ee
so $\Theta\left(f(X^0,X^1)\right)$ is $1$ where the string is located and it vanishes where it is absent.  

Before addressing the ANEC, it is worthwhile to study the energy of the folded string. The energy density above, $T^{00}$,  has the following properties.
In the bulk of the folded string it  is  positive. If we take $X^0-x^0$ to be  large while keeping $X^1-x^1$  finite we get $T^{00}=2/(2\pi \alpha^\prime)$. Namely we get twice the string tension, as expected from a folded string. The energy density  becomes negative as we approach  the folding points, and at  the folding points $T^{00}\to -\infty$. 

The contribution of the negative energy density at the folding region cancels exactly the contribution of the positive energy density in the bulk of the string. Therefore  
for any $X^0$ the total energy associated with this solution vanishes 
\be
E=\int T^{00} \,dx= 0 .
\ee 
This follows from energy conservation and the obvious fact that the folded string did not have energy before it was created.
 
This cancellation supports the claims of \cite{Itzhaki:2018glf} that the driving force behind this solution is a tachyon condensation at the points where the string folds, that is balanced by the folded string tension. The tachyon condensation leads to a negative energy density and the folded string tension gives a positive energy density. 

To calculate the ANEC we have to switch to null coordinates. Defining $u=X^1+X^0$ and $v=X^1-X^0$ we get from (\ref{eq:energy momentum def}) that 
\begin{align}\label{tuu}\nonumber
T_{uu}&=\frac{1}{2 \pi \alpha^\prime} \frac{-e^{-\frac{X^1-x^1}{Q}}}{\sqrt{-1+ \left(2e^{\frac{X^0-x^0}{Q}}-\cosh\left(\frac{X^1 -x^1}{Q}\right)\right)^2}}\,\Theta\left(f(X^0,X^1)\right).\\ 
T_{vv}&=\frac{1}{2 \pi \alpha^\prime} \frac{-e^{\frac{X^1-x^1}{Q}}}{\sqrt{-1+ \left(2e^{\frac{X^0-x^0}{Q}}-\cosh\left(\frac{X^1 -x^1}{Q}\right)\right)^2}}\,\Theta\left(f(X^0,X^1)\right).\\
\nonumber
T_{uv}&=\frac{1}{2 \pi \alpha^\prime} \frac{-\left(2e^{\frac{X^0-x^0}{Q}}-\cosh\left(\frac{X^1-x^1}{Q}\right)\right)}{\sqrt{-1+ \left(2e^{\frac{X^0-x^0}{Q}}-\cosh\left(\frac{X^1-x^1}{Q}\right)\right)^2}}\,\Theta\left(f(X^0,X^1)\right).
\end{align}
Since $T_{uu}$ is negative everywhere along the folded string, the ANEC is negative too
\be\label{r}
\int_{-\infty}^{\infty}du T_{uu}= \frac{1}{2 \pi \alpha^\prime} \left(v-Q \log(4)\right) \Theta\left(-v+Q \log(4)\right),
\ee 
where we set $x^0=x^1=0$.
From (\ref{tuu}) we see  that the main contribution to (\ref{r}) comes from the region where the string folds. 

An interesting feature of  (\ref{r}) is that it becomes more negative with $|v|$ which is a result of the fact that the size of the folded string is growing with time. This plays an important role below.

\section{ The $Q\to 0^+$ limit}

As we are interested in $Q\ll 1$, it is instructive to see what happens when we take $Q\to 0^+$. As we shall see, this limit plays an important role when considering folded strings in the $SL(2,\mathbb{R})_k/U(1)$ BHs. Below we show that a new symmetry emerges in this limit. In the context of the $SL(2,\mathbb{R})_k/U(1)$ BHs this symmetry is essential for the analytic continuation of the FZZ duality \cite{fzz,Kazakov:2000pm}.

There are, however,  some subtleties, associated with taking $Q=0$,  we should discuss before considering the extra symmetry that emerges  when we take $Q\to 0^+$.
Taking $Q\to 0^+$ in the folded string configuration (\ref{s}) we get
\be\label{sq} 
X^{1}=\sigma,~~~~X^{0}=x^{0}+\frac12 \left( \left| (\sigma-x^1)+\tau \right| + \left| (\sigma-x^1)-\tau \right| \right).
\ee
The subtlety is  with the equation that (\ref{sq}) solves. 
If we simply set $Q=0$ in the Virasoro constraints we get, in the $X^{1}=\sigma$ gauge, the following equation
\be\label{vcq} 
1-(\partial_{+}X^0)^2=0,~~~~~~~1-(\partial_{-}X^0)^2=0,
\ee 
which means that 
\be\label{ks} 
\partial_{+} X^0=\pm 1,~~~~~\mbox{and}~~~~~\partial_{-} X^0=\pm 1.
\ee
The solution (\ref{sq}) is obtained by gluing $\partial_{+} X^0=1$ with $\partial_{+} X^0=-1$ (and the same with the right movers). It is easy to see that if we allow such gluing then 
\be\label{qs} 
X^{1}=\sigma,~~~~X^{0}=x^{0}-\frac12 \left( \left| (\sigma-x^1)+\tau \right| + \left| (\sigma-x^1)-\tau \right| \right).
\ee
is also a legitimate solution\footnote{In fact there is a whole family of solutions one obtains by such gluing at several points \cite{Bardeen:1975gx,Bardeen:1976yt,Bars:1994qm}.}.

However, (\ref{qs}) cannot be obtained from (\ref{s}) by taking $Q\to 0$, but it can be obtained by considering negative $Q$ and taking $Q\to 0^{-}$. In other words, only one of the solutions, (\ref{sq}) and (\ref{qs}), survives adding to (\ref{vcq}) an arbitrarily small slope of the dilaton.  The sign of the slope determines which one survives. This suggests that if we simply set $Q=0$, while not specifying which of the limits we are taking,  $Q\to 0^+$ or 
$Q\to 0^-$, there are no folded string solutions at all.

A simple way to see this is the following.\footnote{We thank J. Troost for this argument.} 
Suppose that we simply set $Q=0$, then the Virasoro constraints are (\ref{vcq}). If we take the derivative of (\ref{vcq}) we get 
\be\label{ac} 
\partial^{2}_{+} X^0 \partial_{+} X^{0}=0.
\ee
However
if we glue, say, $\partial_{+} X^0=1$ with 
$\partial_{+} X^0=-1$ at a certain point, then at that point $\partial^{2}_{+} X^0$ blows up, which is not consistent with (\ref{ac}) and the fact that at that point we have either $\partial_{+} X^0=1$ or
$\partial_{+} X^0=-1$. 

This shows that discontinuity in $\partial_{+} X^0$
is not allowed.
The string slope, $Q$, no matter how small,  regularizes this discontinuity to give either (\ref{sq}) or  (\ref{qs}), but not  both.

This is reminiscent of instantons in $U(1)$ non-commutative geometry. Standard $U(1)$ gauge theory  does not have instanton solutions. Turning on non-commutativity one finds solutions \cite{Nekrasov:1998ss}, but, just like in our case, only with orientation that is fixed by the non-commutativity parameter. The size of the $U(1)$ instanton  goes to zero with the non-commutativity parameter, just like the creation region of the folded string scales like  $Q$.

The toy model from the previous section illustrates this too. If we simply set $Q=0$ in (\ref{toy}) then there is no instability, but if we take $Q$ to be positive (negative), then no matter how small $Q$ is, there are positive (negative) and large $E$ modes that are unstable.

Since we consider $Q>0$, the relevant solution for us is (\ref{sq}). This solution covers the entire future wedge of Rindler space. Therefore it is invariant under a boost transformation. To verify that the subtleties discussed  above do not affect this conclusion we calculate the $Q\to 0^{+}$ limit of the energy momentum 
tensor (while keeping $u$ and $v$ finite)
\be 
T_{uu}=\frac{1}{2 \pi \alpha^\prime} v ~\delta(u)~\theta(-v),~~~~T_{vv}=- \frac{1}{2 \pi \alpha^\prime} u~\delta(v)~\theta(u),~~~~T_{uv}=-\frac{1}{2 \pi \alpha^\prime} \theta(-v)~\theta(u),
\ee
that is indeed boost invariant.

Note that the boost generator, $B$, and the total energy, $E$ satisfy the following commutation relation
\be
[B,[B,E]]=E
\ee
which implies that boost invariant states have either infinite or zero energy. The folded string, that is boost invariant in the 
$Q\to 0^{+}$ limit,  has zero energy. The fact that it has the same quantum numbers as the vacuum plays an important role in the BH case we now discussed.

\section{Folded string in the $SL(2,\mathbb{R})_k/U(1)$ BH}

So far we discussed folded strings in a time-like linear dilaton background, where we have an exact solution (\ref{s}). In this section we  wish to argue, without presenting an exact solution, that folded strings are  spontaneously created also behind the horizon of the $SL(2,\mathbb{R})_k/U(1)$ BH and that they violate the ANEC. 

The $SL(2,\mathbb{R})_k/U(1)$ BH background is \cite{Witten:1991yr}
\be\label{SL}
ds^2=-f(r)dt^2+\frac{dr^2}{f(r)},~~~~~\Phi(r)=\Phi_0-Qr
\ee
where $f(r)=1-\mu e^{-2Qr}$.
The local inertial frame around some point $r=r_0$ behind the horizon takes the form of a time-like linear dilaton background
\be\label{lif}
ds^2=-(dX^0)^2+(dX^1)^2, ~~~~\Phi(X^0)=\tilde{\Phi}_0-\tilde{Q} X^0
\ee
where 
 $\tilde{Q}=Q\sqrt{-f(r_0)}$. 

There are curvature corrections to (\ref{lif}) that enter at length scale of the order of $\sqrt{k}\sim 1/Q$. For example, the curvature corrections modify  $T_{+-}$ and  increase the central charge so that, unlike in time-like linear dilaton, it is larger than $2$ (or $3$ in the SUSY case). The curvature corrections also modify $T_{++}$ and $T_{--}$, which control the spontaneous creation and shape of the folded string. However, they do so at scales of the order of $\sqrt{k}\sim 1/Q$, that is much larger than the scale associated with the creation of the folded string, $Q$. 

At first, it seems puzzling that the dilaton gradient and curvature, that are of the same order and have a similar effect on the perturbative string spectrum, have such a different effect on the folded string. The dilaton gradient is, in a sense,  the cause for the folded string's creation, and it affects it at short scales of the order of $Q$, while the curvature enters only at scales of order $1/Q$. The reason is the  following.  What generates the folded string  is the fact that the time-like linear dilaton induces in $T_{++}$ a term that is {\it linear} in $X^0$ 
\be\label{avi}  
\tilde{Q} ~\partial^2_{+} X^0,
\ee
and a similar term in $T_{--}$.
The curvature does not induce linear terms  that can compete with (\ref{avi}) at short distances. Hence the curvature does not prevent the  spontaneous creation of folded strings behind the $SL(2,\mathbb{R})_k/U(1)$ BH horizon. 

The curvature corrections do modify the details of the solution at length scale of the order of $\sqrt{k}$. It seems reasonable to hope that with the help of the $SL(2,\mathbb{R})$ structure, perhaps along the lines of \cite{Yogendran:2018ikf}, one could be able to find the exact solution. As we now argue when $k$ is large, much can be said about the physics associated with the folded string, even without knowing the details of the solution.

In particular, we expect the folded string behind the horizon to violate the ANEC, just like the folded string in the time-like linear dilaton background does. The reason is that the main contribution to the violation of the ANEC comes from the point where the string folds. Away from that point $T_{uu}$ is exponentially suppressed (see (\ref{tuu})). For distances shorter than $\sqrt{k}$ the solution (\ref{s}) is a good approximation for the BH folded string solution. Hence we expect the folded strings in the BH to violate the ANEC.   

We conclude that  the $SL(2,\mathbb{R})_k/U(1)$ BH  is filled with objects that violates the ANEC. The same holds for dynamically formed BHs, since what controls the folded string formation  is  (\ref{op}), that is satisfied behind the horizon.
What is special about the eternal $SL(2,\mathbb{R})_k/U(1)$ BH, is that it is invariant under a boost symmetry  
$
t\to t+C.
$
Creation of a generic folded string behind the horizon will break this symmetry. Interestingly enough, there are folded strings that do not break this symmetry. These are the folded strings that   fill the entire region behind the horizon. In the large $k$ limit at distances smaller than $\sqrt{k}$ from the horizon, such strings are well approximated by  the $Q\to 0^{+}$ limit discussed in the previous section, since $\tilde{Q}\to 0$ when the points where the string folds approach the horizon from the inside.

This suggests that the eternal $SL(2,\mathbb{R})_k/U(1)$ BH is  filled with folded strings that do not break  its symmetries. This could be viewed as  the Lorentzian analog of the FZZ duality.  According to the FZZ duality  the $SL(2,\mathbb{R})_k/U(1)$ cigar geometry is  accompanied with a condensation of a string that wraps  the Euclidean time direction. This condensation does not break the Euclidean time translation symmetry of the cigar. In the FZZ case the condensate of the winding string is correlated with the location of the tip of the cigar. The smaller $g_s$ at the tip of the cigar is, the larger the condensate is. We elaborate more about this in section 6.

There is a possible relation with the SYK model \cite{sy,k}. 
Recently \cite{Saad:2018bqo}  the ramp in $Z(\beta-iT)Z(\beta+iT)$  was proposed to be due to a semi-classical configuration in the SYK model that has a zero action and preserves the symmetries. Here we see, in a different setup, that a  folded string that   fills the entire region behind the horizon has similar properties. Indeed in the Euclidean setup it was shown in \cite{Giveon:2013ica} that the action of the winding  string vanishes. It would be nice to explore this further especially as the ramp is a feature related to the unitarity.

\section{Folded strings vs. the ANEC}

In the context of QFT the ANEC was proved  via information theory \cite{Faulkner:2016mzt} and via causality \cite{Hartman:2016lgu}.  
In this section we discuss the tension between these proofs and the folded strings.
The goal here is not to reconcile the existence of the folded string with these proofs, but rather to emphasis the puzzle, as it is part of the motivation for the conjecture we make in  the next section.
 We start by considering the information theory proof of the ANEC  and then the causality proof.   

\subsection{Information theory vs. folded string}

The basic ingredient in the information theory proof of the ANEC  is the monotonicity of the relative entropy, $S_{rel}$, which, loosely speaking, is saying that it is easier to distinguish between any pure state and the vacuum when we have  access to data in region $A$, than when we have access to data in region $A^{'}$, if $A^{'}\subset A$.

In relativistic QFTs one can consider a setup that also involve the time direction. We describe here the 2D case as it is sufficient for our purposes and  is easier to describe (and draw), but the argument is general \cite{Faulkner:2016mzt}. Suppose that $A$ is a segment at $t=0$ that starts at the origin and ends at $x=L$.    $B(\lambda)$ is a null segment that starts at the origin and ends at $(t=\lambda, x=\lambda)$ with a positive $\lambda$. $A^{'}(\lambda)$ is a straight line that together with $A$ and $B(\lambda)$ forms a triangle (see figure 2).  In a unitary  QFT, for any $\lambda$, the information in $A$ is the same as the information  in $ A^{'}(\lambda)$ and $B(\lambda)$. For example, if we have a string that propagates in space-time, then, since the evolution  is unitary, $A$ contains the same data about the string as  $ A^{'}(\lambda)$ and $B(\lambda)$. This implies that there is more information about the string in $A$ than in $A^{'}(\lambda)$. We lose information by tracing over $B(\lambda)$. As we increase $\lambda$ we lose more information by tracing over $B(\lambda)$ and so  we get 
\be\label{o} 
\frac{d S_{rel}(\lambda)}{d\lambda}\leq 0,
\ee
where $S_{rel}(\lambda)$ is the relative entropy in $A^{'}(\lambda)$.

To get the ANEC one takes $L\to\infty$, which gives the right wedges of Rindler space for which the modular Hamiltonian, that enters the definition of the relative entropy, is 
$ H =2\pi\int_{0}^{\infty} du u T_{uu}$. See \cite{Faulkner:2016mzt} for details.
The upshot is that (\ref{o}), which is very intuitive, implies the ANEC. Since the folded string violates the ANEC, it must contradict (\ref{o}). There are two aspects to this contradiction we wish to  discuss.

The first is that, unlike a standard string that propagates in a unitary fashion in space-time, the folded string is spontaneously created at a certain time and space. Therefore for a given folded string there are $A$, $B(\lambda)$ and $A^{'}(\lambda)$ such that the folded string simply does not exist at $A$, but does exist at $A^{'}(\lambda)$ (see figure 2). So there exists a  $\lambda$ such that  $S_{rel}(\lambda)>S_{rel}(0)=0$ which contradicts (\ref{o}). 
The reason is that a folded string solution that is put by hand at a certain time and place is not unitary. Unitarity implies that the seed of this particular folded string should be present at earlier times, including on $A$. In that sense the situation here is similar to the one discussed in \cite{Gao:2016bin}.  There to violate the ANEC a certain squeezed operator is inserted at $t=0$. This insertion creates a state that is propagated forwards,  but not backwards in time. So the state at $t>0$ is not a unitary evolution of the state at  $t<0$. 

In the BH case (see figure 3), the situation is slightly better. At $A$ we already have the information that a BH will be formed and it will be filled with folded strings. However, we still do not know the exact state of these folded strings. A detailed  understanding of the folded string creation 
mechanism is needed to figure out this state.

The second aspect in which the folded string contradicts (\ref{o}) has no analogy in \cite{Gao:2016bin}.  As we increase $\lambda$ it is getting {\it easier } to distinguish the folded string from the vacuum. Figure 2 is useful to illustrate this. Normally as we increase $\lambda$ we lose more information when tracing over $B(\lambda)$. Here however, (\ref{r}) implies that the total flux that is crossing $A^{''}$ is larger than the flux that is crossing $A^{'}$. Hence it is easier to distinguish the folded string from the vacuum when tracing over $B^{'}$, than when tracing over $B$. This too contradicts (\ref{o}). This contradiction appears to be more severe since it cannot be resolved by understanding the details of the creation mechanism; once the folded string is created it gets larger, which makes it easier to distinguish it from the vacuum.

\begin{figure}\label{fh2}
\begin{center}
\includegraphics[scale=0.40]{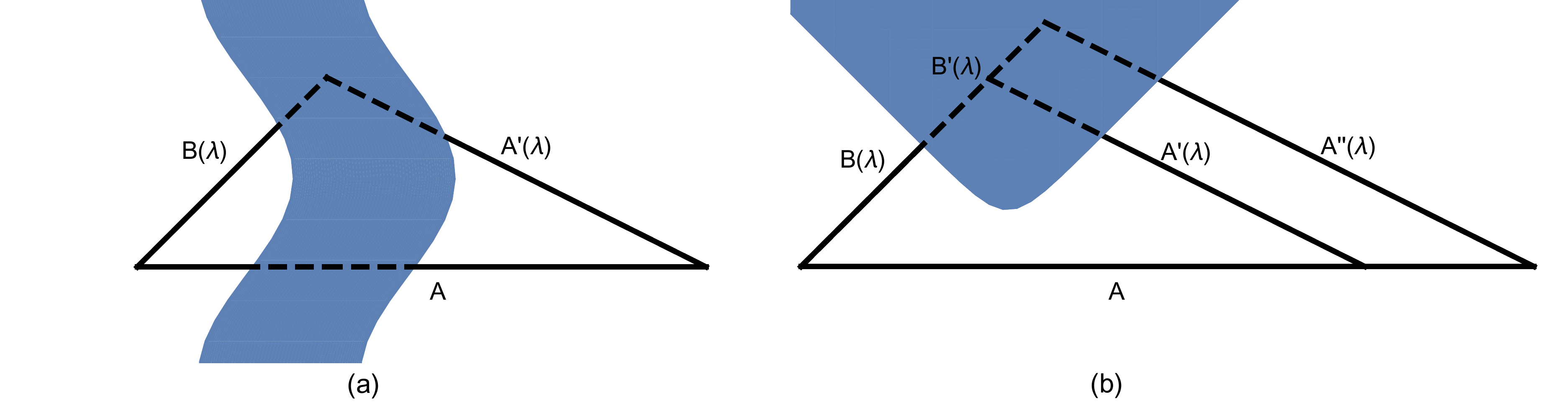}\vspace{-2mm}
\caption{In standard situations, like a unitary evolution of a string (a), the information contained in $A$ is larger or equal to the one in $A^{'}(\lambda)$. This is not the case with the folded string (b). Using the information at $A$ we cannot distinguish the folded string from the vacuum, as it was not yet created. Moreover, since according to (\ref{r}) the flux that crosses $A^{''}$ is larger than the flux that crosses $A^{'}$, the relative entropy is increasing. }
\label{SBHpen}
\end{center}
\end{figure}

This leads to the following comment about  the Quantum Null Energy Condition (QNEC) \cite{Bousso:2015mna,Bousso:2015wca} and the folded string.
In standard cases the ANEC can be derived from the QNEC. Therefore it is natural to expect the folded string to violate the QNEC too. However, as we now show this is not necessarily the case. The main point is as following. In standard situations the fact  that $S_{rel}(\lambda)\geq 0$ for every $\lambda$
 implies that 
\be\label{oha} 
\lim_{\lambda\to\infty} \frac{d S_{rel}(\lambda)}{d\lambda}=0.
\ee  
 This is crucial for deriving the ANEC from  the QNEC. 
In terms of the relative entropy,  the QNEC is the statement that in QFT there is a condition, that does not hold in general quantum systems,  that
\be\label{a}
\frac{d^2 S_{rel}}{d\lambda^2}\geq 0.
\ee
To show that (\ref{o}) follows from (\ref{a}) we  simply integrate (\ref{a})
\be\label{av}
\int_{\lambda}^{\infty} d\tilde{\lambda}\frac{d^2 S_{rel}}{d\tilde{\lambda}^2}= \frac{d S_{rel}}{d\tilde{\lambda}} (\infty)-\frac{d S_{rel}}{d\tilde{\lambda}} (\lambda),
\ee  
which implies  (\ref{o}) when (\ref{oha}) and
(\ref{a}) hold.

The folded string, however, does not fit the standard case:  according to (\ref{r}) the flux keeps on growing indefinitely with $\lambda$. Thus   we do not expect (\ref{oha}) to hold. Hence it is possible that the folded string violates the ANEC, but not the QNEC.   
Needless to say that it would be interesting to understand this better.

\subsection{Causality and the folded string}

\begin{figure}\label{fh2}
\begin{center}
\includegraphics[scale=0.30]{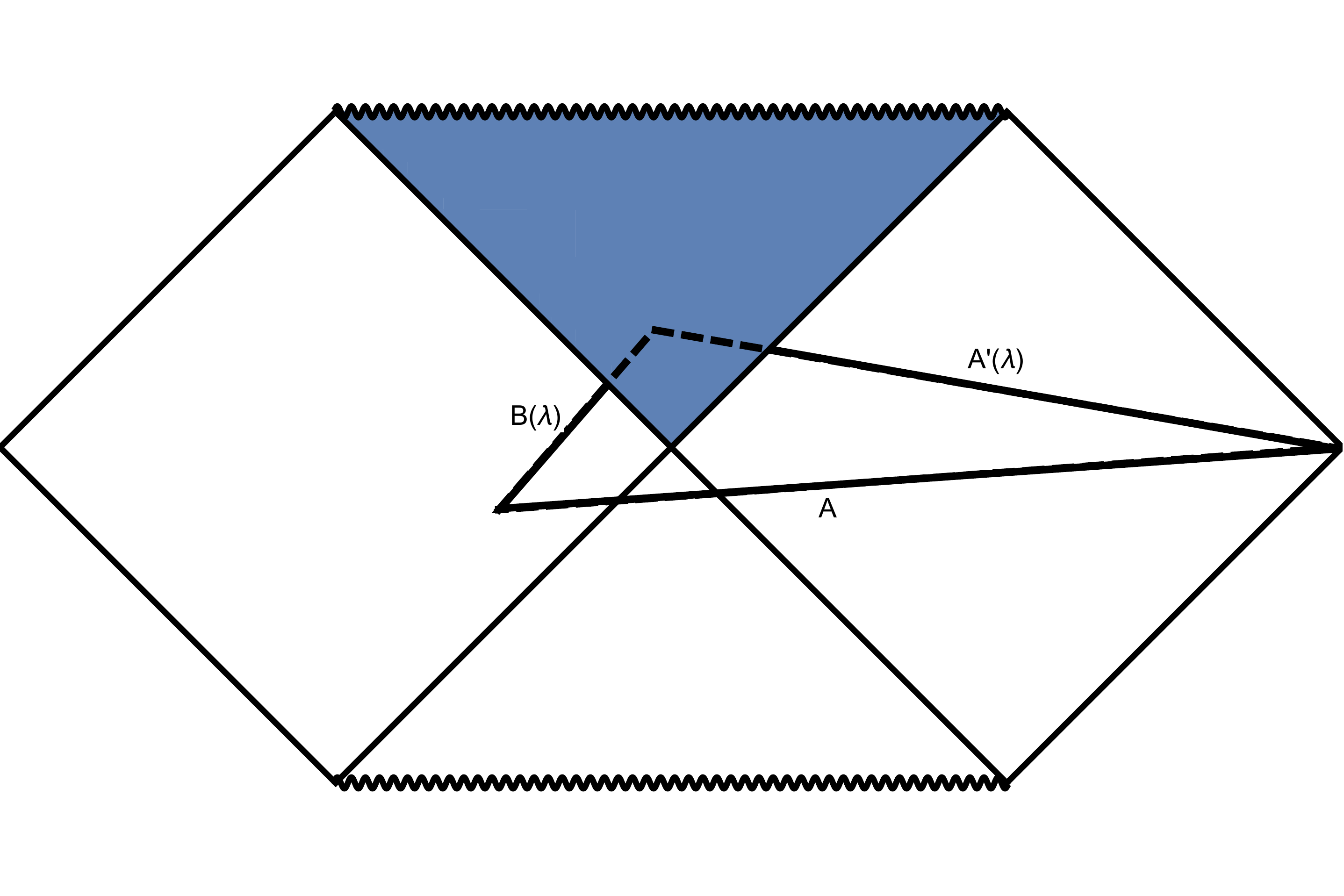}\vspace{-2mm}
\caption{In the BH case the situation is a bit better. Already at $A$ we know that  folded strings will be formed since we know the initial conditions  that lead to the BH formation. We do not know, however, the details of the folded strings.}
\label{SBHpen}
\end{center}
\end{figure}

The fact that there is  tension between causality and the folded string is clear from the solution, as the points where the string folds move faster than light. In this subsection we discuss different aspects of causality that are related to the proof of the ANEC.

The causality proof of the ANEC can be phrased  in the following way. Consider a null ray that follows the geodesics $v=v_0$. If the  ANEC is violated, then the null ray will be pushed to earlier null time,  $v_1<v_0$. In unitary QFT this was shown using a careful study of the analytic properties of the 4-point function \cite{Hartman:2016lgu}. 

In theories that involve gravity, like string theory, we have
\be\label{shift} 
\Delta v \sim G_N \int du T_{uu}, 
\ee
where $G_N\sim g_s^2$ is the Newton constant and $\Delta v=v_1-v_0$. This together with (\ref{r}), imply that the folded string violates causality, as $\Delta v<0$. Moreover, from (\ref{r}) we see that 
\be\label{on}
\Delta v \sim G_N v
\ee
which means that by taking large $v$  we can have a macroscopic $\Delta v$.

With the help of two folded strings and a mirror, a macroscopic closed time-like curve can be formed.  This is illustrated in figure 4.
One can argue against this construction    that it assumes we can choose the location of the folded strings,   and adjust the location of the mirror accordingly, while in fact the folded strings are created spontaneously and it is not clear how to control their location. Still it is disturbing that if we randomly choose a location for the mirror there is a chance  a macroscopic  closed time-like curve is formed.

\section{A conjecture}

In the previous section we saw that there is a clear tension between the fact that the folded string violates the ANEC and 
 unitarity and causality.  Being a consistent theory, string theory is expected to resolve this tension. In that respect there is a big difference between the time-like linear dilaton and the $SL(2,\mathbb{R})_k/U(1)$
 BH. 

\begin{figure}\label{fh2}
\begin{center}
\includegraphics[scale=0.40]{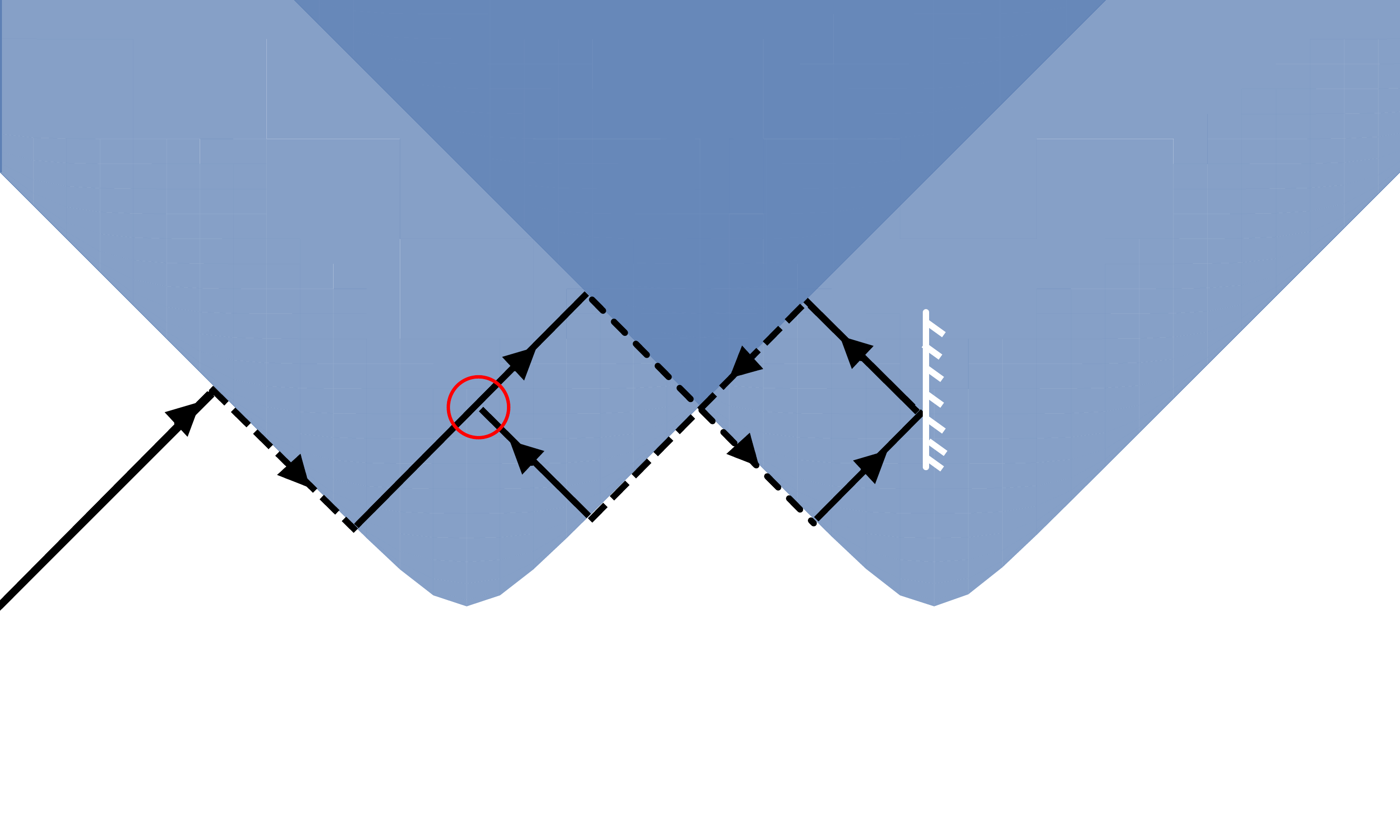}\vspace{-17mm}
\caption{With the help of two folded strings and a mirror, a closed time-like curve is formed. The trajectory of the light ray is represented by a solid line and the  effect of (\ref{on}) by a dashed line.  }
\label{SBHpen}
\end{center}
\end{figure}

The time-like linear dilaton, by itself,  is not a consistent background of superstring theory at the quantum level . For example, its central charge is not critical.  We merely used it as a simple setup in which  the classical solution of the folded string can be found exactly. As a quantum theory it must be supplemented with an additional CFT so that the full theory is critical. In the case of a large $Q$ this is easy to do, but for small $Q$ things are more subtle. We could add a space-like linear dilaton (times $R^8$), but then we end up with a null linear dilaton background that admits different physics altogether. We could add a cigar geometry (times $R^7$) and focus on  the tip of the cigar where the dilaton is approximately a constant. This, however, could mix the time-like linear dilaton effects with complications  associated with the tip of the cigar \cite{Giveon:2013ica,Giveon:2015cma,Ben-Israel:2015mda}.  

The $SL(2,\mathbb{R})_k/U(1)$
 BH background, on the other hand, is obtained in 10D superstring as the near horizon limit of k near extremal NS5-branes \cite{Maldacena:1997cg}.
Hence it is a consistent background that should lead to consistent physics.    
Since we do not have a detailed understanding of the mechanism that creates the folded strings, we cannot show from first principle  how  string theory resolve the causality and unitarity  issues in the BH case. We can, however, make a   conjecture.

The  conjecture is that the number of folded strings that are spontaneously created  in the BH is,
\be\label{no}
N\sim\frac{1}{g_s^2},
\ee
which implies together with (\ref{on})  that
\be\label{ck}
\Delta v\sim v
\ee
and similarly $\Delta u\sim u$. 

More precisely we conjecture that $N$ is such that there is an equality in (\ref{ck})
\be\label{ef}
\Delta v= v,~~~\mbox{and}~~~\Delta u =u.
\ee 
This means 
 that the BH interior is cloaked by the folded strings. A null ray that follows $v=v_0$ will be pushed all the way to the horizon bifurcation at $v=u=0$ and continue on the other side. In other words, the folded strings prevent information from falling into the BH.

There are several, related,  arguments that support this conjecture:\\
$\bullet$ What drives the creation of the folded string is the 
time-like linear dilaton. The back reaction of the folded string is expected to suppress the time-like linear dilaton. Since the tension of the string is of order $1$ and  $G_N=g_s^2$, the number of folded string needed to suppress an order one effect in the background is (\ref{no}).\\
$\bullet$ The conjecture prevents null rays from penetrating the folded string. Hence the closed  time-like curve, discussed in the previous section, cannot be formed.   \\
$\bullet$ The tension with information theory is resolved since  the folded string backreaction prevents information from falling into the BH. \\
$\bullet$ The 2D BH entropy is $1/g_s^2$ which, according to the conjecture, is the number of folded strings. This suggests that, in effect, the BH is replaced by the folded strings. This is very much related to the next point.\\
$\bullet$ The conjecture seems to  fit neatly with the FZZ duality.  The FZZ duality implies that  the $SL(2,\mathbb{R})_k/U(1)$ cigar CFT is dual to the CFT of a cylinder with a Sine-Liouville condensation. From a stringy point of view the Sine-Liouville is  a winding string mode. In the large $k$ limit it was shown in \cite{Giveon:2015cma} that the way to think about the duality is that at low energies (compared to $\sqrt{k}$) the cigar is a good description and the Sine-Liouville description takes over at the deep UV.

\begin{figure}\label{fh2}\vspace{-15mm}
\begin{center}
\includegraphics[scale=0.40]{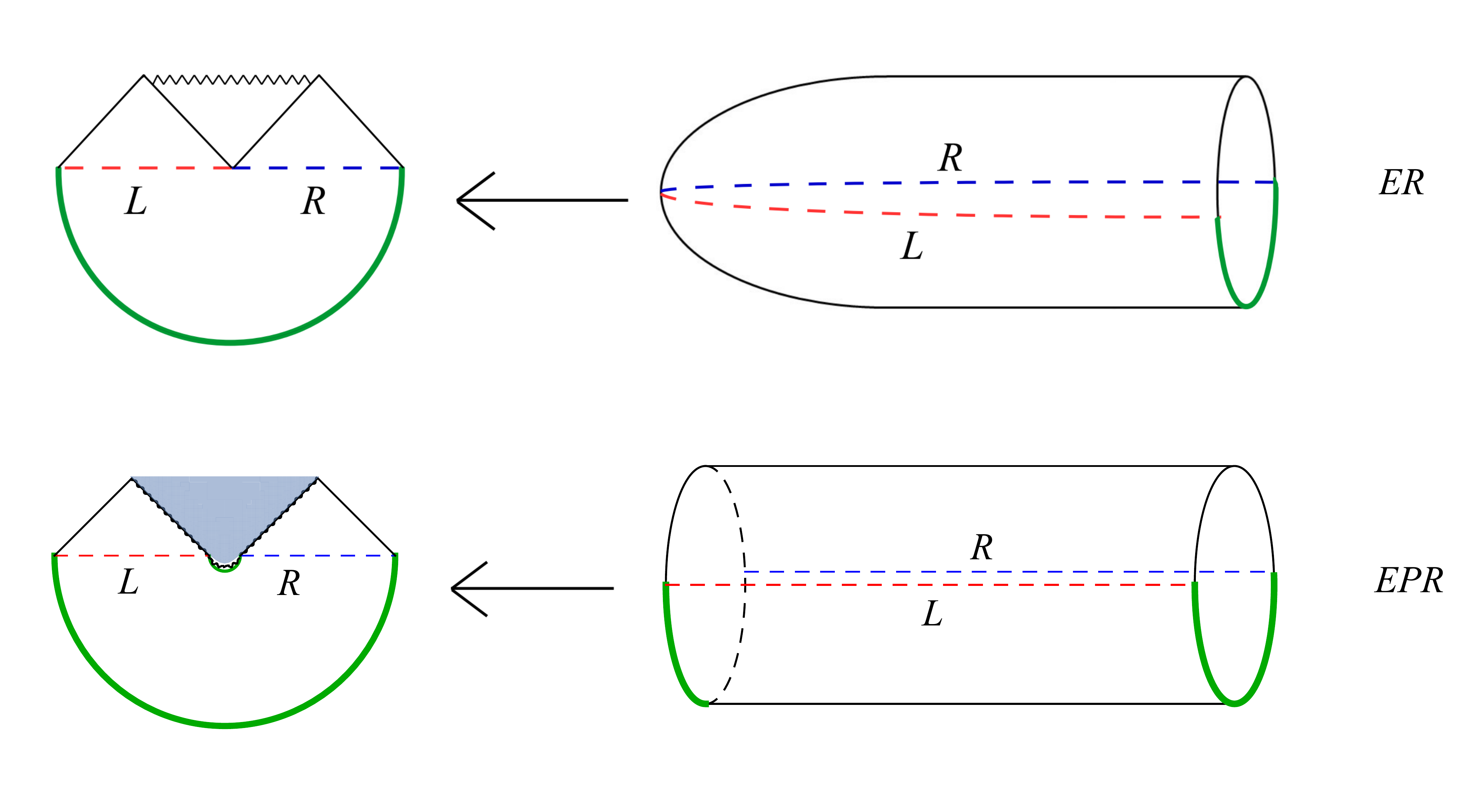}\vspace{-10mm}
\caption{The FZZ duality combined with the Hartle-Hawking wave function and the conjecture suggests a relation that is somewhat in the spirit of the ER=EPR proposal. The ER side is an effective  picture valid for an observer that can probe both the Euclidean and the Lorentzian configuration only with low energy modes. The EPR side is the fine-grained picture  an observer with access to all energies sees. The fact that  the folded strings violate the ANEC is key. Their backreaction eliminates the BH interior while entangling  the left and right sides of the horizon.  }
\label{SBHpen}
\end{center}
\end{figure}

The Hartle-Hawking wave function \cite{Hartle:1976tp} then suggests a duality that is somewhat in the spirit of the ER=EPR  of \cite{Maldacena:2013xja}. An observer at infinity that has access only to low energies will conclude that the eternal BH solution is a good approximation. This is the ER side of the duality. However, an observer that can probe the BH also with high energy modes will conclude that the region just behind the horizon is singular \cite{Ben-Israel:2017zyi,Itzhaki:2018rld}. This is the EPR side of the duality that is related to the cylinder with the Sine-Liouville condensation. Unlike in the case of the cigar, when we, following \cite{Hartle:1976tp}, cut the cylinder  we get two lines that do not meet. As a result the analytic continuation gives two  asymptotic regions that are not connected (see figure 5). These regions, however, are entangled by the analytic continuation of the winding string. This is exactly what the folded strings are doing according to the conjecture. Their backreaction  eliminates the BH interior and entangles the information between the two horizons.  

Note that a signal from, say,  the left wedge that hits the folded strings is stuck at the right horizon but it does not make it to the right infinity. This means that the conjecture does not imply traversable wormhole.\\ 
$\bullet$ By now there are various reasons to think that the Hawking particles should be on-shell close to the horizon for the information to get out \cite{Itzhaki:1996jt,Braunstein:2009my,Mathur:2009hf,Almheiri:2012rt,Marolf:2013dba,Polchinski:2015cea}. Namely they form a firewall \cite{Almheiri:2012rt}.

A firewall can prevent the information from falling into the BH after it was formed. However, information can fall into the BH long before the firewall was formed, even if the firewall is formed  together with the BH and not after the page time \cite{Page:1993wv}. To see this consider the situation in figure 6. Several soft photons that carry some information are emitted at some time from infinity towards the origin. At a later time an energetic null spherically symmetric shell falls to form a BH. Even if we assume that a firewall is formed as soon as possible, it is not able to prevent the early photons from falling to the singularity. If their information is contained also in the radiation then we have the usual problematic situation in which information is copied. To resolve this we have to resort to ideas in the spirit of the BH complementarity \cite{tHooft:1990fkf,Susskind:1993if}. 

This means that a firewall is not enough to recover all the information, unless it is extended into the entire BH interior. This is exactly what the conjecture implies.      

\section{Summary}

\begin{figure}\label{fh2}
\begin{center}
\includegraphics[scale=0.35]{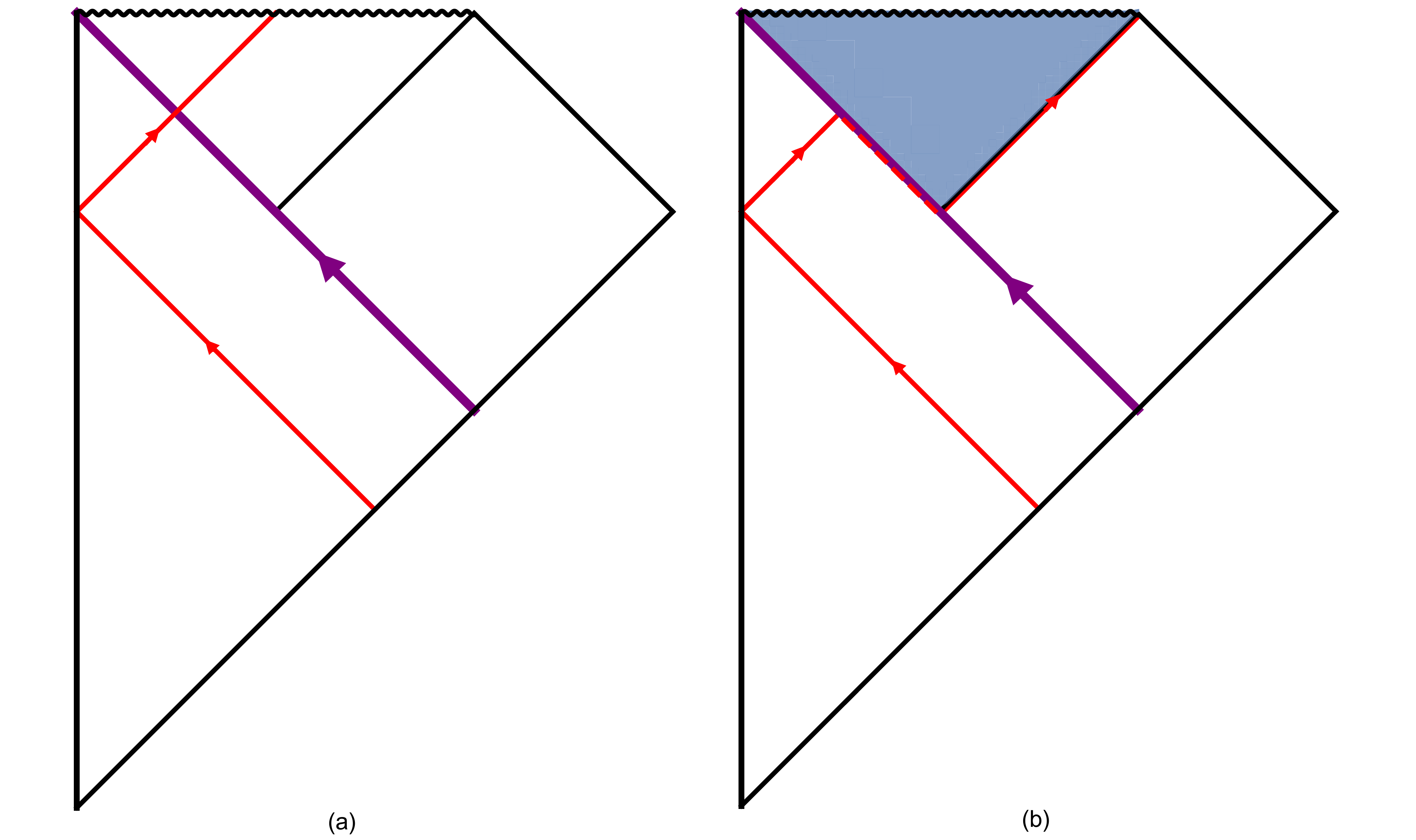}\vspace{-2mm}
\caption{Firewall is not enough: The information carried by early soft photons (red line)  is lost at the singularity even if a firewall (black line) is formed as early as possible (a). According to the conjecture, the number of folded strings that are created is such that their backreaction pushes the photons to the horizon (b).  } 
\label{SBHpen}
\end{center}
\end{figure}

The main point of the paper is to show that the folded strings of \cite{Itzhaki:2018glf}, that are spontaneously created behind the horizon of the $SL(2,\mathbb{R})_k/U(1)$ BH, violate the ANEC macroscopically. This suggests that the region behind the horizon is in a different phase than the region outside the BH - a phase in which the ANEC is violated. 

To understand the details of this phase we should, at least, be able to calculate, from first principle, the creation rate of these folded string. Unfortunately as far as we can see such a calculation involves understanding closed string field theory in time dependent situations which is way beyond our abilities. Hence we had to settle for  a conjecture about the nature of this phase. The conjecture is that the number of folded strings that are created is such that their backreaction prevents information from falling into the BH. Some indirect evidence for the conjecture was presented. 

\section*{Acknowledgments}

We thank L. Liram and J. Troost for discussions.
This work  is supported in part by the I-CORE Program of the Planning and Budgeting Committee and the Israel Science Foundation (Center No. 1937/12), and by a center of excellence supported by the Israel Science Foundation (grant number 1989/14).

\end{document}